\documentclass[prd,preprint,superscriptaddress,a4paper,showpacs,showkeys,11pt,nofootinbib]{revtex4}
\usepackage{graphicx}
\usepackage{graphics}
\usepackage{bm}
\usepackage{multirow}
\usepackage{tabularx}
\usepackage{hyperref}
\usepackage{commath}
\usepackage{epstopdf}
\usepackage[T1]{fontenc}
\usepackage[latin9]{inputenc}
\usepackage{geometry}
\usepackage{soul}
\usepackage{amsmath,amssymb,amsfonts}
\usepackage{dcolumn}
\usepackage{color}

\title{Heavy quark production}

\def\be{\begin{equation}}
\def\ee{\end{equation}}

\providecommand{\ee}{e$^+$e$^-$}

\makeatother
\newcommand{\pom}{$\mathbb P\,$}

\newcolumntype{Y}{>{\centering\arraybackslash}X}

\begin{document}
\title{Single charged Higgs pair production in exclusive processes at the LHC}

\author{Laura {\sc Duarte}}
\email{l.duarte@unesp.br}
\affiliation{Institute of Physics and Mathematics, Federal University of Pelotas, \\
  Postal Code 354,  96010-900, Pelotas, RS, Brazil}

\author{Victor P. {\sc Gon\c{c}alves}}
\email{barros@ufpel.edu.br}
\affiliation{Institute of Physics and Mathematics, Federal University of Pelotas, \\
  Postal Code 354,  96010-900, Pelotas, RS, Brazil}

\author{Daniel E. {\sc Martins}}
\email{daniel.ernani@ifj.edu.pl}
\affiliation{The Henryk Niewodniczanski Institute of Nuclear Physics (IFJ)\\ Polish Academy of Sciences (PAN), 31-342, Krakow, Poland
}

\author{T\'essio B. de {\sc Melo}}
\email{tessiomelo@institutosaphir.cl}
\affiliation{Universidad Andr\'{e}s Bello, Facultad de Ciencias Exactas, \\ Departamento de Ciencias F\'{i}sicas-Center for Theoretical and Experimental Particle Physics, \\
Fern\'{a}ndez Concha 700, Santiago, Chile\\}
\affiliation{Millennium Institute for Subatomic Physics at the High-Energy Frontier, SAPHIR, Chile}

\begin{abstract}
The production of a single charged Higgs boson pair by photon - photon interactions  in $pp$ collisions at the LHC is investigated in this exploratory study. We focus on the exclusive production, which is characterized by intact protons and two - rapidity gaps in the final state, and  assume the type - I two - Higgs - doublet model, which still allows a light charged Higgs. Assuming the 
leptonic $H^{\pm}\rightarrow [\tau\nu_{\tau}]$ decay mode, we derive predictions for the transverse momentum, rapidity and invariant mass distributions of the $\tau^+ \tau^-$ pair for different values of the charged Higgs mass. The contribution of different background processes are also estimated. Our results indicate that the contribution of the exclusive $H^+ H^-$ production for the 
$[\tau^{+}\nu_{\tau}][\tau^{-} \nu_{\tau}]$ final state is non - negligible and can, in principle, be used to searching for a light charged Higgs.
\end{abstract}

\maketitle

\section{Introduction}

The discovery of the Higgs boson at the LHC in 2012 is one the great triumphs of the Particle Physics, and represents the completion of the Standard Model (SM) \cite{ATLAS:2012yve, CMS:2012qbp}. However, there are still many unanswered questions that suggest that the SM is an effective low - energy realization of a more complete and fundamental theory. Several scenarios for the beyond the Standard Model 
(BSM) physics predict the presence of extra physical Higgs boson states, which has  motivated the searching for these 
additional states in various production and decay channels over a 
wide range of kinematical regimes at LEP, Tevatron and now at the LHC \cite{Bahl:2021str, Moretti:2016jkp, Arhrib:2018bxc, LEP(20010), Guchait:2001pi, Akeroyd:2019mvt, ATLAS:2020zms, CMS:2018rmh, ATLAS:2018gfm, ATLAS:2021upq, CMS:2019bfg, CMS:2019rlz, Arhrib:2018ewj}. 
As these studies have yield negative, constraints have been placed on the 
associated masses and branching ratios of different decay channels. In particular, the existence of a light charged Higgs boson, with mass below the top quark mass, is still allowed only in a restrict number of BSM models \cite{Akeroyd:2016ssd, Arhrib:2017wmo, Drees:1998pw, Chakraborti:2021bpy, Aoki:2011wd, Davoudiasl:2014mqa}.

The two-Higgs-Doublet model (2HDM) is one the simplest BSM frameworks that predict  charged Higgs bosons \cite{Branco:2011iw, Aoki:2009ha}. In this model,  an additional complex doublet is added   and its Higgs sector involves five scalars:  CP-even neutral $h$ and $H$, CP - odd neutral $A$, and a pair of charged Higgs $H^{\pm}$. Four distinct interaction modes arise when a $Z_2$ symmetry is introduced to prevent the Flavor Changing Neutral Currents (FCNC) at the tree level. Current experimental measurements imply  that a light $H^{\pm}$ can only be accommodated in type-I 2HDM  and type-X 2HDM \cite{Arhrib:2017wmo, Cheung:2022ndq, Arhrib:2021xmc}. In this paper, we will concentrate our analysis in the type - I 2HDM, where one of doublets couples to all fermions, and a light charged Higgs with mass below 100 GeV is still allowed. 

Over the last decades, extensive phenomenological studies on the $H^{\pm}$ production in $e^+ e^-$, $\gamma \gamma$,
 $e p$ and $pp$ collisions have been performed assuming the type - I 2HDM. In particular, a comprehensive analysis of the single charged Higgs pair production at the LHC has been recently performed in Ref.\cite{Cheung:2022ndq}. Such a study considers the charged Higgs pair production in inelastic processes, where both the incident protons breakup, and the dominant subprocesses are initiated by the quarks and gluons present in the proton wave function. Moreover, they have explored the entire parameter of type - I 2HDM and obtained the phenomenologically viable parameters. The authors also have demonstrated that the significance for the $pp \rightarrow H^+ H^- \rightarrow [\tau \nu] [\tau \nu]$ channel is large and that a future experimental analysis of this final state is a promising way for searching and discover  the single charged Higgs boson. 
 
\begin{figure}[t]
\includegraphics[width=0.425\textwidth,height=5.8cm]{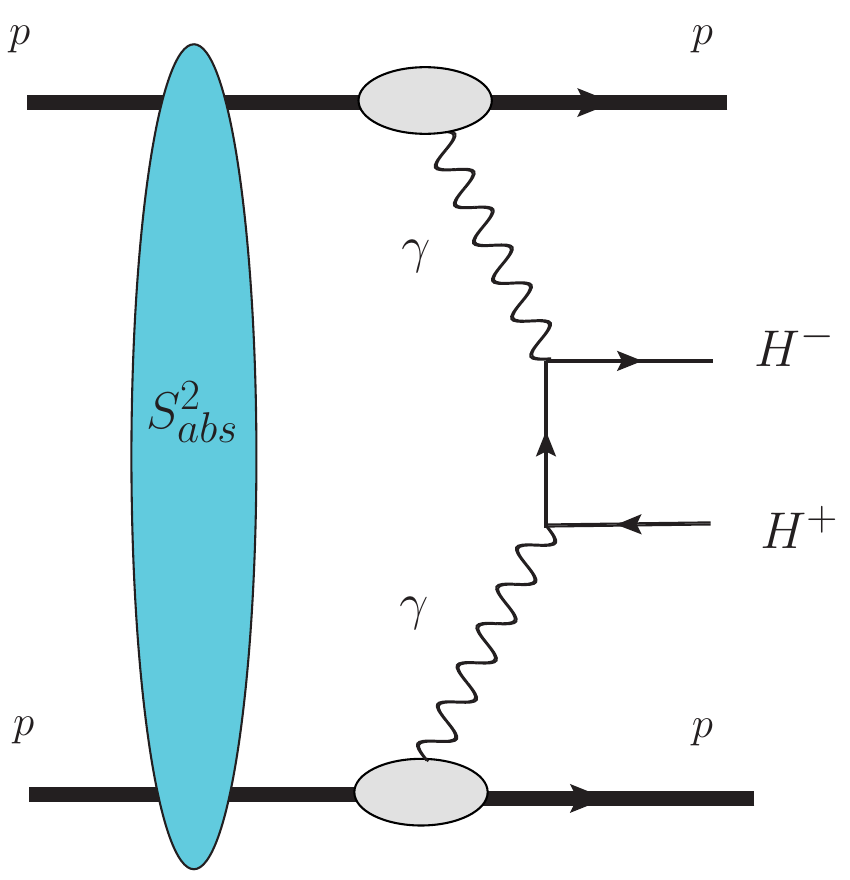}
\caption{Single charged Higgs pair production by $\gamma \gamma$ interactions in $pp$ collisions at the LHC.}
\label{Fig:Diagram}
\end{figure}

 The main goal in this paper is to extend the analysis performed in Ref. \cite{Cheung:2022ndq} for elastic processes, where the two incident protons remain intact in the final state. In order to protons remain intact, the $H^+ H^-$ pair should be produced by the interaction of color singlet objects, which can be a photon $\gamma$, a $Z$ boson or a Pomeron \pom, which is a  color singlet particle with partonic structure.  
As the dominant channel is the $\gamma \gamma \rightarrow H^+ H^-$ subprocess \cite{Lebiedowicz:2015cea}, we will focus on the process represented in Fig. \ref{Fig:Diagram}, which is usually denoted the  exclusive $H^+ H^-$ production, since only the $H^+ H^-$ pair is present in the final state.   Our analysis is strongly motivated by the recent study performed in Ref.~\cite{Duarte:2022xpm}, where we have considered the exclusive production of a pair of doubly charged Higgs and demonstrated that such a process can be used to search for signatures of the type
II seesaw mechanism and to obtain lower mass bounds on $H^{\pm \pm}$. We will focus on the $pp \rightarrow p \otimes H^+ H^- \otimes p  \rightarrow p  \otimes [\tau \nu] [\tau \nu]  \otimes p$ channel, where $\otimes$ represents the presence of  rapidity gaps in the final state, and will estimate  the total cross-section and associated differential distributions on the transverse momentum, rapidity and invariant mass of the $\tau \tau$ pair, considering $pp$ collisions at $\sqrt{s} = 14$ TeV and different values of the $H^{\pm}$ mass. A comparison with the predictions associated to the potential SM backgrounds will also be performed.

This paper is organized as follows. In the next section, we present a brief review of the type - I 2HDM and of the formalism used  for the treatment of the $H^+ H^-$ pair production by photon-induced interactions in $pp$ collisions. In Section \ref{sec:res} we discuss the backgrounds considered in our analysis and present our predictions for the invariant mass,
transverse momentum and rapidity distributions, as well as for the total cross-section for the $H^+ H^-$ pair production in $\gamma \gamma$ interactions. Finally, in Section \ref{sec:sum}  we summarize our main conclusions.

\section{Formalism}

\subsection{The type-I 2HDM}

The charged scalar bosons appear in several
extensions of the SM \cite{Branco:2011iw,Craig:2012vn, Eriksson:2008cx, Cheung:2022ndq, Eberhardt:2013uba}.  
A two-Higgs-doublet model (2HDM) is a simple extension of the SM by
introducing an additional $SU(2)_L$ Higgs doublet, which predicts three neutral Higgs bosons
and a pair of charged Higgs bosons $H^{\pm}$. 
To realize this model, the fermion content are the same as in the SM, while the scalar sector is extended with two 
$SU(2) _L$ scalar doublets $\Phi _1$ and $\Phi _2$, which have weak hypercharge $Y = 1$:
\begin{equation}
\Phi _i = \begin{pmatrix} \phi _i ^+ \\ \frac{v _i + \rho _i + i \eta _i}{\sqrt{2}} \end{pmatrix} , \ \ \ \ \ \ \ \ \ \ \  i = 1, 2 \ ,
\end{equation}
with $v _1$ and $v _2$ the vacuum expectation values (VEVs) of $\Phi _1$ and $\Phi _2$, respectively. $v _1$ and $v _2$ satisfy the relation $v = \sqrt{v _1 ^2 + v _2 ^2} = 246$ GeV, in order to successfully generate the electroweak symmetry breaking, and the ratio of $v _2$ and $v _1$ defines the mixing angle $\beta$ as $\tan \beta = \frac{v _2}{v _1}$.
In what follows,
we use the simplified notation of $s_x = \sin x$, $c_x = \cos x$ and $t_x = \tan x$.

The physical mass eigenstates are given by 
\begin{align}
\begin{pmatrix}  G^\pm \\ 
H^\pm \end{pmatrix} = R(\beta)\begin{pmatrix}  \phi_1^\pm \\ 
\phi_2^\pm \end{pmatrix}, 
&&
\begin{pmatrix}  G \\ 
A \end{pmatrix} = R(\beta)\begin{pmatrix}  \eta_1 \\ 
\eta_2 \end{pmatrix},
&&
    \begin{pmatrix}  H\\ 
h \end{pmatrix} = R(\beta)\begin{pmatrix}  \rho_1 \\ 
\rho_2 \end{pmatrix}.
\end{align}
Here $G^\pm$ and $G$ are the Nambu-Goldstone bosons that are
eaten as the longitudinal components of the massive gauge
bosons. The rotation matrix is
\begin{eqnarray}
  R(\theta)  = \begin{pmatrix} \cos\theta & \sin\theta\\ 
-\sin\theta & \cos\theta \end{pmatrix}.
\end{eqnarray}


General 2HDMs are troubled by the presence of flavor changing neutral currents (FCNC) at tree level, as the gauge symmetries allow both doublets to couple to all fermions.
In order to prevent large FCNC, additional symmetries may be employed, in order to forbid some of the offending couplings. A popular choice is to impose a discrete $Z _2$ symmetry\footnote{Continuous Abelian symmetries, either global or gauged, have also been considered in this context. See, e.g., \cite{Campos:2017dgc, Davidson:2009ha, Camargo:2018uzw, Ko:2012hd, Ko:2013zsa}.}, under which $\Phi _1 \to - \Phi _1$.
The fermion $Z _2$ parities are such that each type of fermion couples to only one of the doublets.
The different possibilities to fulfill this condition give rise to the well known four types of 2HDMs \cite{Branco:2011iw}: In the type-I model, only the doublet $\Phi _2$ couples to all the fermions so all the quarks and charged leptons get their
masses from the VEV of $\Phi _2$ (ie. $v_2$); in the type-II, $\Phi _1$ couples to charged leptons and down-type quarks and $\Phi _2$ to up-type quarks; in the type-X, the charged leptons couple to $\Phi _1$ and all the quarks couple to $\Phi _2$, while in the type-Y model, $\Phi _1$ couples to down-type quarks and $\Phi _2$ to up-type quarks and charged leptons. 


The most general scalar potential which is CP-conserving and invariant under the $Z _2$ symmetry (up to the soft-breaking term proportional to $m _{12} ^2$) is given by:
\begin{equation}
\label{potential}
\begin{split}
V = & m _{11} ^2 \Phi _1 ^\dagger \Phi _1 + m _{22} ^2 \Phi _2 ^\dagger \Phi _2 - m _{12} ^2 \left( \Phi _1 ^\dagger \Phi _2 + \Phi _2 ^\dagger \Phi _1 \right) + \frac{\lambda _1}{2} \left( \Phi _1 ^\dagger \Phi _1 \right) ^2 + \frac{\lambda _2}{2} \left( \Phi _2 ^\dagger \Phi _2 \right) ^2 \\
& + \lambda _3 \Phi _1 ^\dagger \Phi _1 \Phi _2 ^\dagger \Phi _2 + \lambda _4 \Phi _1 ^\dagger \Phi _2 \Phi _2 ^\dagger \Phi _1 + \frac{\lambda _5}{2} \left[ \left( \Phi _1 ^\dagger \Phi _2 \right) ^2 + \left( \Phi _2 ^\dagger \Phi _1 \right) ^2 \right] .
\end{split}
\end{equation}
After spontaneous symmetry breaking, five physical scalars arise, three neutral $h$, $H$, $A$ and a pair of charged scalars $H ^\pm$. The remaining three scalars become the longitudinal components of the massive $W ^{\pm}$ and $Z$ gauge bosons. As the CP symmetry is conserved, the neutral CP-even and CP-odd states do not mix, which means they can be diagonalized separately. The mixing angle for the CP-even sector is denoted by $\alpha$, while $\beta$ is the mixing angle for the CP-odd sector, as well as for the charged sector.

The eight parameters $m_{ij}^2$ and $\lambda_1 - \lambda_5$ in the potential \eqref{potential} are replaced by the VEV $v$, the mixing angles $\alpha$ and
$\beta$ the scalar masses $M_h$, $M_H$, $M_A$ and $M_H^{\pm}$, and the soft $Z_2$ breaking parameter
$M^2 = m_{12}^2/s_{\beta}c_{\beta}$. In particular, the quartic coupling constants are given as \cite{Cheung:2022ndq}
\begin{eqnarray}
&& \lambda_1 = \frac{1}{v^2}[(s_{\beta-\alpha}-c_{\beta-\alpha}t_{\beta})^2+M_H^2(s_{\beta-\alpha} t_{\beta}+c_{\beta-\alpha})^2+M^2 t_{\beta}^2]\nonumber\\   
&&\lambda_2 = \frac{1}{v^2}\left[M_h^2\left(s_{\beta-\alpha}+\frac{c_{\beta-\alpha}}{t_{\beta}}\right)^2-\frac{M^2}{t_{\beta}^2}+M_{H}^2\left(\frac{s_{\beta-\alpha}}{t_{\beta}}-c_{\beta}\right)^2\right] \nonumber\\
&&\lambda_3 = \frac{1}{v^2}\left[(M_h^2-M_H^2)\left\{s_{\beta-\alpha}^2-s_{\beta-\alpha} c_{\beta-\alpha}\left(t_{\beta}-\frac{1}{t_{\beta}}\right)-c_{\beta-\alpha}^2\right\}+2M_{H^{\pm}}^2-M^2\right]\nonumber\\
&&\lambda_4 = \frac{1}{v^2}[M^2+M_{A}^2-2M_{H^{\pm}}^2]\nonumber\\
&&\lambda_5 = \frac{1}{v^2}[M^2-M_{A}^2].
\end{eqnarray}

The general Yukawa Lagrangian with two scalar doublets is
\begin{equation}
\begin{split}
\label{yuk_lag_gener}
- \mathcal{L} _Y & = y ^{1d} \bar{Q} _L \Phi _1 d _R + y ^{1u} \bar{Q} _L \Tilde{\Phi} _1 u _R + y ^{1e} \bar{L} _L \Phi _1 e _R \\
& + y ^{2d} \bar{Q} _L \Phi _2 d _R + y ^{2u} \bar{Q} _L \Tilde{\Phi} _2 u _R + y ^{2e} \bar{L} _L \Phi _2 e _R,
\end{split}
\end{equation}
where $Q_L^T = (u_L, d_L)$, $L_L^T= (\nu_l, l_l)$ and $y^{ (1, 2)(d,u,e)}$ are $3\times 3$ matrices
in family space. After imposing the $Z _2$ symmetry, some of these terms are forbidden.
The production of a charged Higgs particle, depending on its mass with respect to the top quark, can be divided into
light ($M_{H^\pm} \ll M_t$), intermediate ($M_{H^\pm} \sim M_t$) and heavy
($M_{H^\pm} \gg M_t$) scenarios \cite{Degrande:2016hyf, Flechl:2014wfa, Cheung:2022ndq}.
Current searches impose stringent constraints on the mass of $H ^\pm$ depending on the 2HDM type. 
For instance, the charged Higgs boson in type-II and type-Y is tightly constrained
to be as heavy as $M_H^\pm \gtrsim 800$ GeV due to the measurements of the inclusive weak radiative $B$ meson decay into $s\gamma$\cite{Misiak:2015xwa, Haller:2018nnx}. Only type-I and type-X can accommodate a light $H^{\pm}$. 
As we are interested in studying relatively light charged scalars, with a mass $\sim 100$ GeV, we will focus here in the type-I 2HDM, in which such small masses are allowed provided that $\tan \beta$ is not too small. 
In this case, all the Yukawa couplings of $H^{\pm}$ are inversely proportional
to $\tan\beta$ and the decay
branching ratios into a fermion pair are proportional to the fermion mass. 
Restricting to the type-I model, the Yukawa Lagrangian \eqref{yuk_lag_gener} in the physical basis takes the form:
\begin{equation}
\begin{split}
- \mathcal{L} _Y ^{\text{Type-I}} = & \frac{1}{t_\beta} \left\{ \sum _f \frac{m _f}{v} \left( \frac{c_ \alpha}{c_\beta} h \bar{f} f + \frac{s_\alpha}{c_ \beta} H \bar{f} f \right) - i A (\bar{u} \gamma _5 u - \bar{d} \gamma _5 d - \bar{l} \gamma _5 l) \right. \\
& \left. + \left[ \frac{\sqrt{2} V _{ud}}{v} H ^+ \bar{u} ( m _u P _L - m _d P _R ) d - \frac{\sqrt{2} m _l}{v} H ^+ \bar{\nu} _L l _R + h. c. \right] \right\}
\end{split}
\end{equation}
with $f = u, d, l$. Notice that the coupling of the Higgs bosons to the fermions become suppressed when $t_\beta > 1$.
The interaction of the scalars with the gauge bosons is given by,
\begin{equation}
\begin{split}
\mathcal{L} _{\text{gauge}} & = \left( g m _W W _\mu ^\dagger W ^\mu + \frac{1}{2} g _Z m _Z Z _\mu Z ^\mu \right) [ s_{\beta - \alpha} h + c_ {\beta - \alpha} H ] \\
& + \frac{g}{2} i [ W _\mu ^+ [ c_{\beta - \alpha} h - s_{\beta - \alpha} H ] \overleftrightarrow{\partial ^\mu} H ^- - h.c.] - \frac{g}{2} [ W _\mu ^+ H ^- \overleftrightarrow{\partial ^\mu} A + h.c. ] \\ 
& + i \left[ e A _\mu + \frac{g _Z}{2} ( \sin ^2 \theta _W - \cos ^2 \theta _W ) Z _\mu \right] H ^+ \overleftrightarrow{\partial ^\mu} H ^- + \frac{g _Z}{2} Z _\mu [ \cos (\beta - \alpha) A \overleftrightarrow{\partial ^\mu} h - \sin (\beta - \alpha) A \overleftrightarrow{\partial ^\mu} H ] ,
\end{split}
\end{equation}
where $g _Z = g / \cos \theta _W$ and  $f \overleftrightarrow{\partial ^\mu} g \equiv ( f \partial ^\mu g - g \partial ^\mu f ) $.



We concentrate on the usual scenario where $h$ is the SM-like $125$ GeV Higgs boson, with being $H$ the heavier neutral scalar. The pseudoscalar $A$ can be either heavier or lighter than $h$.
The parameter ranges are determined by the small $H^\pm$ mass scenario we are interested in, and also by the several experimental constraints: 
LEP experiments \cite{LEP(20010)} have given limits on the mass of the charged Higgs boson
in 2HDM from the charged Higgs searches in Drell-Yan
events, $e^+ e^- \to Z \gamma \to H^+ H^-$, excluding $m_{H^\pm}  \lesssim 80$ GeV (Type II) and $m_{H^\pm} \lesssim 72.5$ GeV (Type I) at 95\% confidence level. Among the constraints from B meson decays
(flavor physics constraints), the $B \to X_s\gamma$ decay \cite{HFLAV:2014fzu} puts a
very strong constraint on Type II and Type Y 2HDM, excluding $m_H^{\pm}\lesssim 580$ GeV and almost independently of $t_\beta$. For
Type I and Type X, the $B \to X_s\gamma$ constraint is sensitive only
for low $t_{\beta}$. 
%

In Ref. \cite{Cheung:2022ndq}, the authors have performed a comprehensive study and explored the entire parameter space for type - I 2DHM, deriving the current viable parameters for these models, taking into account the current theoretical and experimental constraints. In our analysis, we will make use of the results obtained in Ref. \cite{Cheung:2022ndq} and, in particular, we  will estimate the exclusive cross-section for the parameters associated to the benchmark point 1 defined in Table III of that reference, in which $M_H =138.6$ GeV, $M_{A} =120.7$ GeV, $\tan \beta  = 16.8$, $\sin (\beta - \alpha)  = 0.975$, $m_{12}^{2} = 1089.7$ GeV$^{2}$ and $m_h$ being the mass of the observed Higgs boson. For $M_{H^{\pm}}$, three distinct values will be considered: $M_{H^{\pm}} = 80,\, 100$ and 140 GeV.


\subsection{Single charged Higgs pair production by $\gamma \gamma$ interactions}

Assuming the validity of the equivalent photon approximation (EPA) \cite{epa}, the total  cross section for the $H^+ H^-$ production by $\gamma \gamma$  interactions in $pp$ collisions can be factorized in terms of the equivalent flux of
photons into the proton projectiles and the $\gamma \gamma \rightarrow H^+ H^-$ cross section, as follows
\begin{eqnarray}
\sigma(pp \rightarrow p \otimes H^{+} H^{-} \otimes p) = S^2_{abs}  \int dx_1 \int dx_2 \, \gamma^{el}_1(x_1) \cdot \gamma^{el}_2(x_2) \cdot \hat{\sigma}(\gamma \gamma \rightarrow H^{+} H^{-}) \,\,,
\label{fotfot}
\end{eqnarray}
where  $x$ is the fraction of the proton energy carried by the photon
and $\gamma^{el}(x)$ is the elastic equivalent photon distribution of the proton. 
The general expression for the elastic photon flux of the proton has been derived in Ref.  \cite{kniehl} and is given by 
\begin{eqnarray}
\gamma^{el} (x) = - \frac{\alpha}{2\pi} \int_{-\infty}^{-\frac{m^2x^2}{1-x}} \frac{dt}{t}\left\{\left[2\left(\frac{1}{x}-1\right) + \frac{2m^2x}{t}\right]H_1(t) + xG_M^2(t)\right\}\,\,,
\label{elastic}
\end{eqnarray}
where $t = q^2$ is the momentum transfer squared of the photon,
\begin{eqnarray}
 H_1(t) \equiv \frac{G_E^2(t) + \tau G_M^2(t) }{1 + \tau}
\end{eqnarray}
with $\tau \equiv -t/m^2$, $m$ being the proton mass, and where $G_E$ and
$G_M$ are the Sachs elastic form factors. In our analysis,  we will use the photon
flux derived in Ref.~\cite{epa}, where an analytical expression is presented. 
Moreover,  $S^2_{abs}$ is the absorptive factor, which takes into account of additional soft interactions between incident protons which leads to an extra production of particles that
destroy the rapidity gaps in the final state \cite{bjorken}. In our study, we will assume that $S^2_{abs} = 1$, which is a reasonable approximation since the contribution of the soft
interactions is expected to be small in $\gamma \gamma$ interactions due to the long range of the
electromagnetic interaction (For a more detailed discussion see, e.g., Ref. \cite{Lebiedowicz:2015cea}). The cross section for the  $\gamma \gamma \rightarrow H^+ H^-$ subprocess, $\hat{\sigma}(\gamma \gamma \rightarrow H^{+} H^{-})$, will be estimated using the type - I 2DHM and the events associated with the signal will be generated by MadGraph~5 \cite{madgraph}.

A comment is in order here. The cross sections for the exclusive production of a given final state system, in general, two orders of magnitude smaller than for the production in inelastic proton - proton collisions, where both incident
protons break up and a large number of particles is produced in addition to the final state. It turned out that the analysis of e.g. the $H^+ H^-$ production in inelastic collisions, generally involve serious backgrounds, thus making the search for new physics a hard task. In contrast,  exclusive processes have  smaller backgrounds and are characterized by a very clean  final state,  identified by the 
presence of two rapidity gaps, i.e. two regions devoid of hadronic activity separating the intact very
forward protons from the central system. Such exclusive events  can be clearly distinguished from the inelastic one by detecting the scattered protons in
spectrometers placed in the very forward region close to the beam pipe, such as the ATLAS Forward Proton
detector (AFP) \cite{Adamczyk:2015cjy,Tasevsky:2015xya} and the CMS--Totem Precision Proton Spectrometer
(CT--PPS) \cite{Albrow:2014lrm}, and selecting events with two rapidity gaps in the central detector.
For a more detailed discussion about the separation of exclusive processes we refer the interested reader to the recent studies performed in Refs. \cite{Goncalves:2020saa,Martins:2022dfg}.

\section{Results}
\label{sec:res}


\begin{table}[t]
\begin{center}
\scalebox{0.85}{
\begin{tabularx}{\textwidth}{|c|*{3}{Y|} c|c|}
\hline
\hline 
$pp$ @ 14 TeV & \multicolumn{3}{c|}{\bf Signal} & \multicolumn{2}{c|}{\bf Backgrounds}   \\     
\hline 
\hline
 & \multicolumn{3}{c|}{$pp \rightarrow p H^{+}H^{-}p \rightarrow p [\tau^{+}\nu_{\tau}][\tau^{-}\bar{\nu}_{\tau}]p$}   & $ pp \rightarrow p W^{+}W^{-} p \rightarrow p [\tau^{+}\nu_{\tau}][\tau^{-} \nu_{\tau}]p$ & 
 $pp \rightarrow p \tau^{-}\tau^{+} p$    \\     
    \hline
$M_{H^{\pm}}$ [GeV]&   80    & 110    & 140    &  -  &  -               \\
\hline
\hline
   $\sigma$(fb)& 0.20   & 0.14  & 0.10     &1.21 &235500   \\
    \hline  
    \hline      
\end{tabularx}}
\end{center}
\caption{Predictions at the generation level for the total cross-sections for the single charged Higgs pair production  via photon-photon interactions in $pp$ collisions at $\sqrt{s} = 14$ TeV,  derived assuming the Type - I 2HDM and different values for the mass $M_{H^{\pm}}$. Results for the main backgrounds are also presented.} 
\label{tab:tau}
\end{table}

In what follows we will present our results for the single charged Higgs pair production in exclusive processes considering $pp$ collisions at $\sqrt{s} = 14$ TeV. In our analysis, we  assume that the $H^{+}H^{-}$ system decays leptonically,
$H^{+}H^{-}\rightarrow [\tau^{+}\nu_{\tau}][\tau^{-}\bar{\nu}_{\tau}] $ and analyze two distinct experimental scenarios: 
\begin{itemize}
    \item {\bf Scenario I:} the $\tau^{+}\tau^{-}$ pair is produced at central rapidities ($ -2.0 \le y(\tau \tau) \le + 2.0$) and  both forward protons are tagged, which is the configuration that can be studied by the ATLAS and CMS Collaborations. We require both forward protons to be detected by Forward Proton Detectors (FPDs) and we will we assume an efficient reconstruction in the range
$0.012 < \xi_{1,2} < 0.15$, where $\xi_{1,2} = 1 - p_{z 1,2}/E_{\rm beam}$ is the
fractional proton momentum loss on either side of the interaction point
(side 1 or 2) and $p_{z 1}$ is the longitudinal momentum of the scattered proton on the side 1.
This, in principle, allows one to measure masses of the central
system by the missing mass method, $m_X\!=\!\sqrt{\xi_1\xi_2s}$, starting from
about 160~GeV.
    \item {\bf Scenario II:} the $\tau^{+}\tau^{-}$ pair is produced at forward rapidities ($ + 2.0 \le y(\tau \tau) \le + 4.5$), but the protons in the final state are not tagged, which is the case that can be analyzed by the LHCb Collaboration.
\end{itemize}
The signal is assumed to be the $pp \rightarrow p H^{+}H^{-}p \rightarrow p [\tau^{+}\nu_{\tau}][\tau^{-}\bar{\nu}_{\tau}]p$ process, which will be generated using the   MadGraph~5 \cite{Christensen:2008py, madgraph}. For the background we will consider the photon - induced processes 
 $pp \rightarrow p \otimes  W^+ W^- \otimes p \rightarrow p \otimes  [\tau^{+}\nu_{\tau}][\tau^{-}\bar{\nu}_{\tau}] \otimes p$ and $pp \rightarrow p \otimes  \tau^+\tau^- \otimes p$. All these backgrounds are also generated by MadGraph~5 \cite{madgraph}.  

\begin{figure}[t]
\includegraphics[width=0.425\textwidth,height=4.6cm]{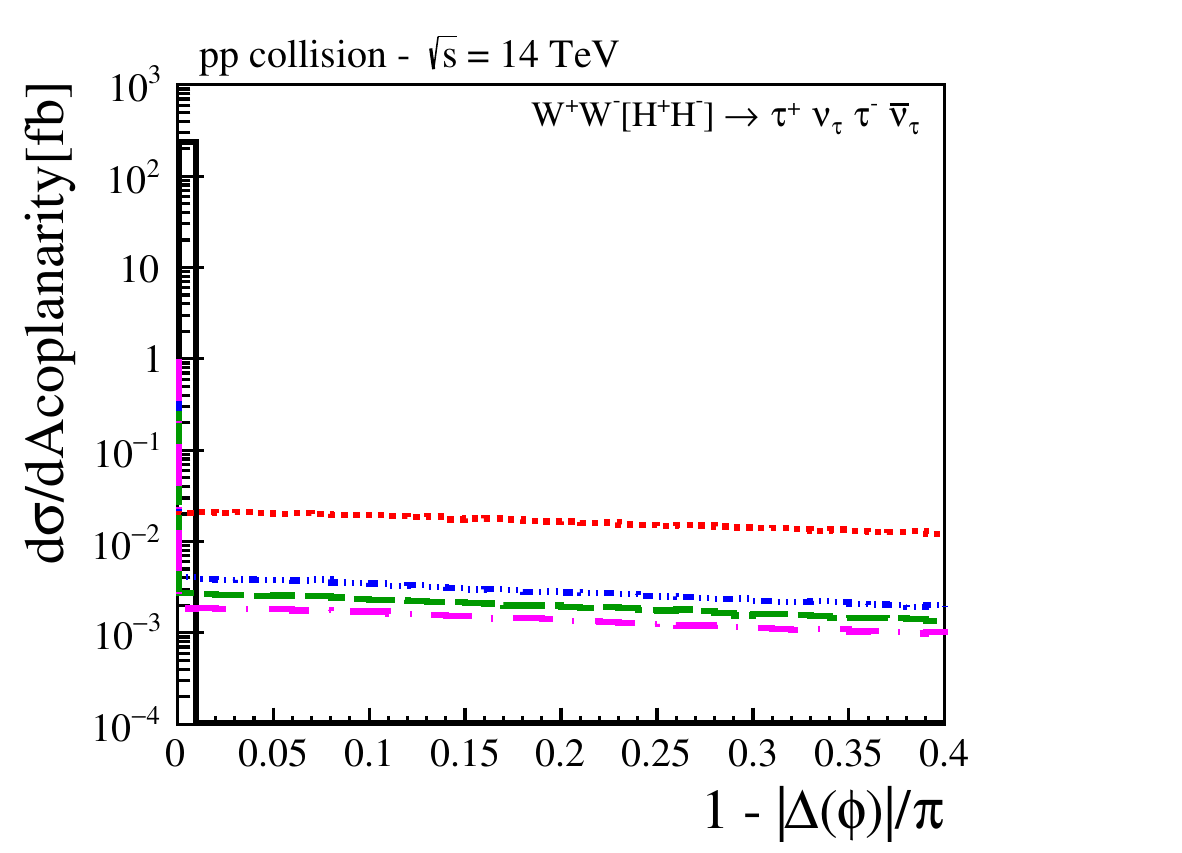} 
\includegraphics[width=0.425\textwidth,height=4.6cm]{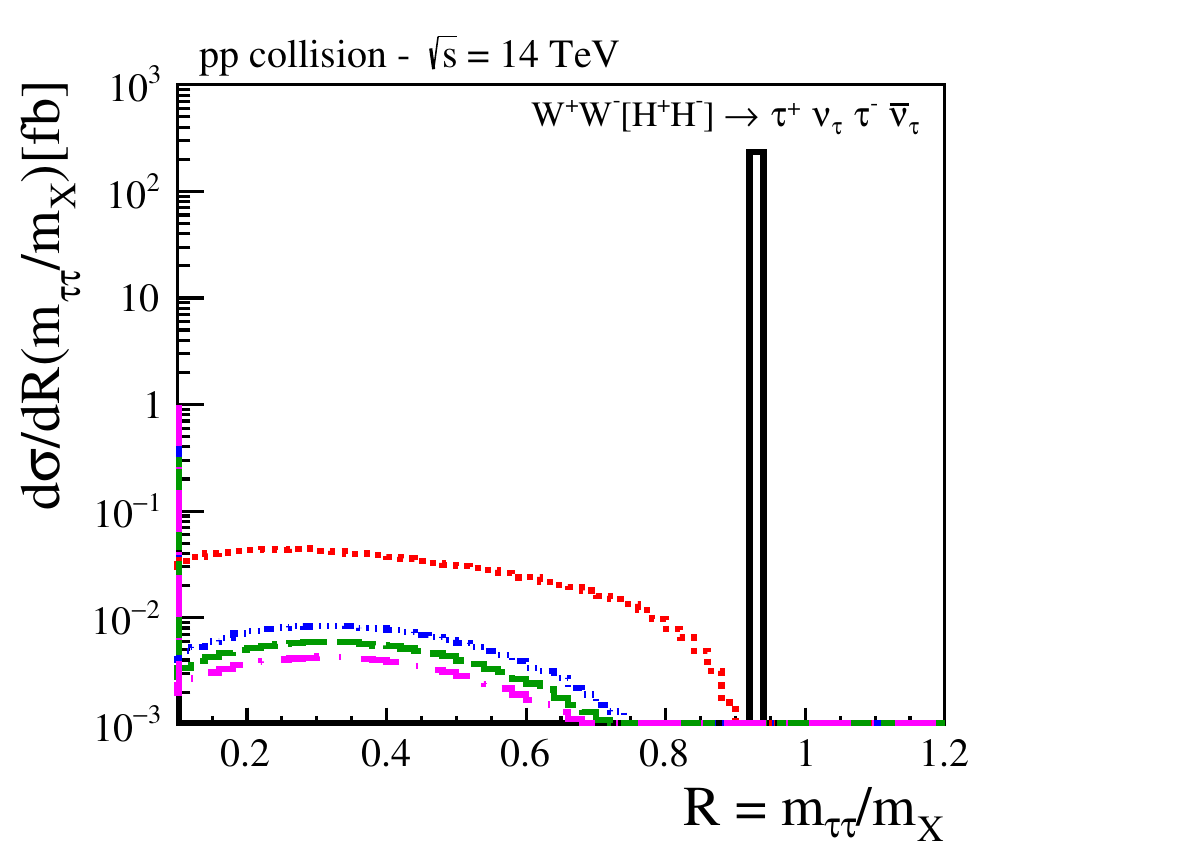} \\

\caption{Predictions for the acoplanarity distribution (left panel) and for the dependence on the ratio $R = m_{\tau \tau}/m_X$ (right panel) for the cross sections associated with the signal and backgrounds. The results for the signal have been derived assuming different values for the mass $M_{H^{\pm}}$.}
\label{Fig:acop}
\end{figure}

In Table \ref{tab:tau} we present our predictions for the total cross-sections associated with the signal and backgrounds considered in our analysis. The results for the signal have been derived assuming the type - I 2HDM for the benchmark point 1 \cite{Cheung:2022ndq}, discussed in the previous Section,  and different values for the mass  $M_{H^{\pm}}$. The predictions for the signal at the generation level are similar to that derived in Ref. \cite{Lebiedowicz:2015cea}.  One has that the $pp\rightarrow p \otimes  \tau^+\tau^- \otimes p$ process dominates the production of a 
$\tau^+\tau^-$ pair. However, such a process is characterized by a distinct topology in comparison to the other channels, where neutrinos (not seen by the detector) are also produced. In particular, this process is characterized by a pair with small acoplanarity  [$ \equiv 1 -(\Delta \phi/\pi)$], where $\Delta \phi$ is the angle between the $\tau$ particles, since the pair is dominantly produced back - to - back. Moreover, it is also characterized by an invariant mass $m_{\tau \tau}$ almost identical to the measured mass of the central system $m_X$, since other particle are not produced in addition to the $\tau^+\tau^-$ pair. Such expectations are confirmed by the results presented in Fig. 
\ref{Fig:acop}, where we show the acoplanarity distribution associated to the different channels (left panel) and the cross-sections as a function of the ratio $R = m_{\tau \tau}/m_X$ (right panel). 
These results suggest that the contribution of the $pp\rightarrow p \otimes  \tau^+\tau^- \otimes p$ process can be strongly suppressed by assuming cuts on the acoplanarity and  on the ratio $R$. In what follows, we will explore such expectations. 


\begin{figure}[t]
\includegraphics[width=0.425\textwidth,height=5.8cm]{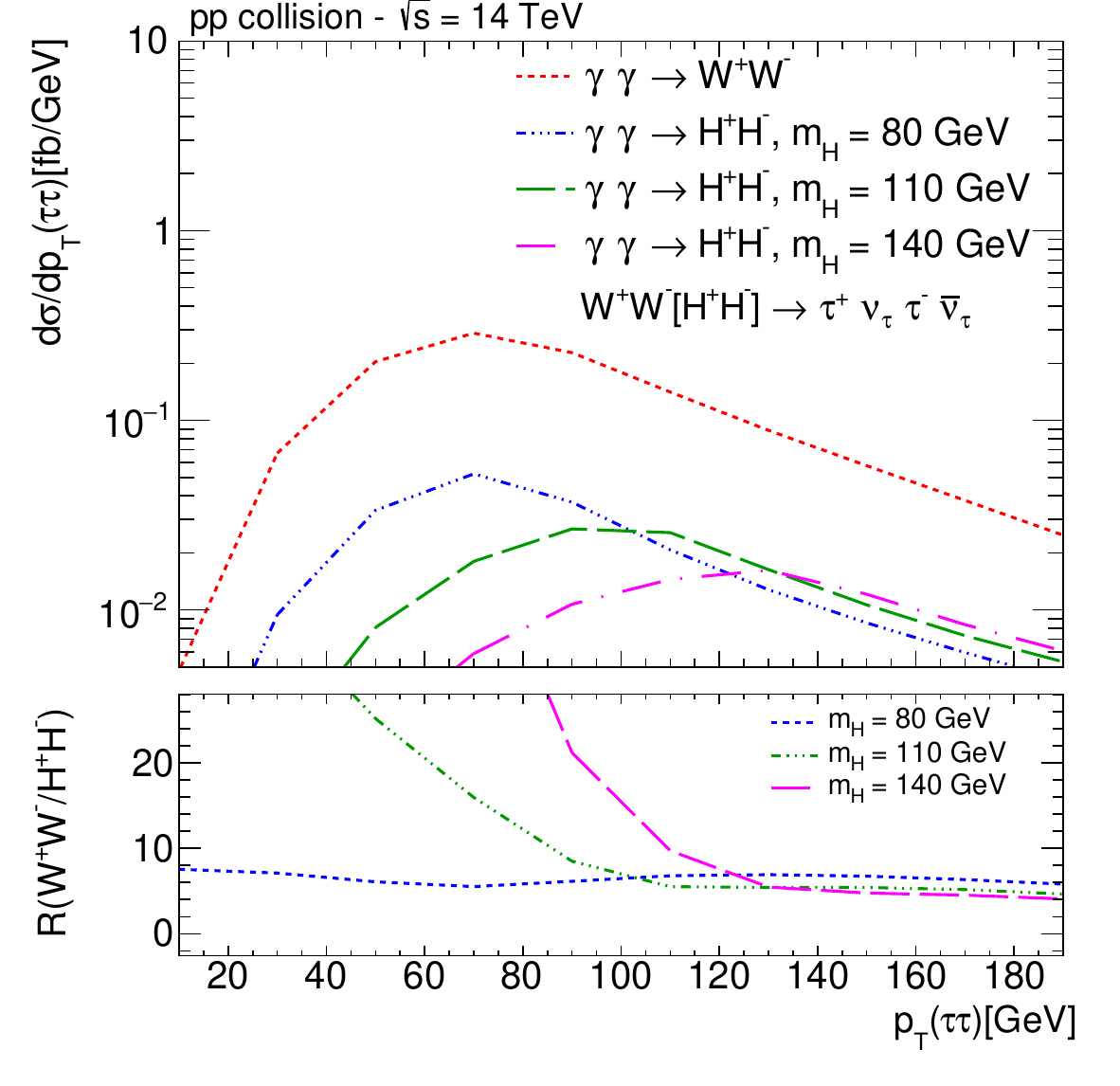} 
\includegraphics[width=0.425\textwidth,height=5.8cm]{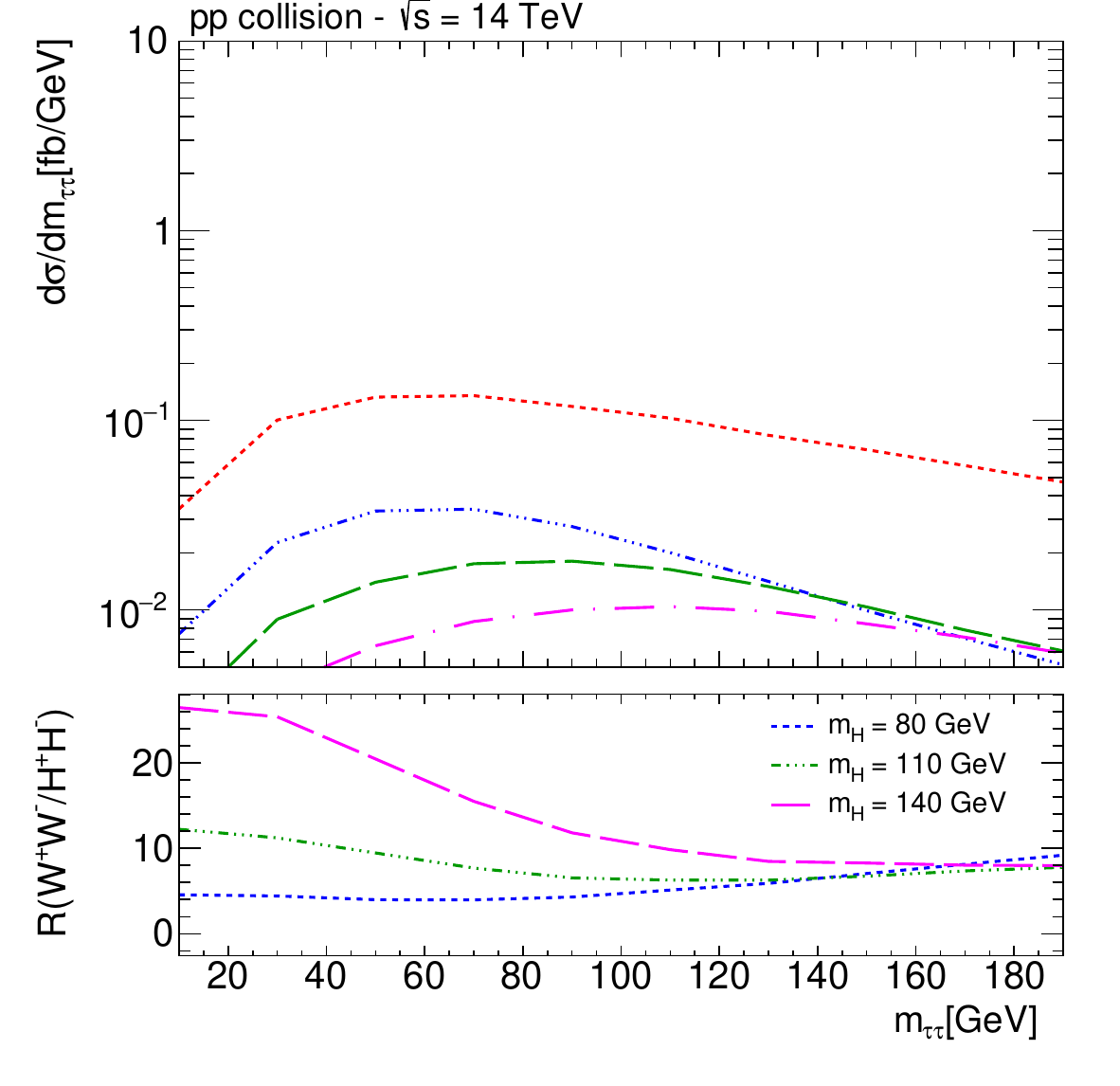} \\
\includegraphics[width=0.425\textwidth,height=5.8cm]{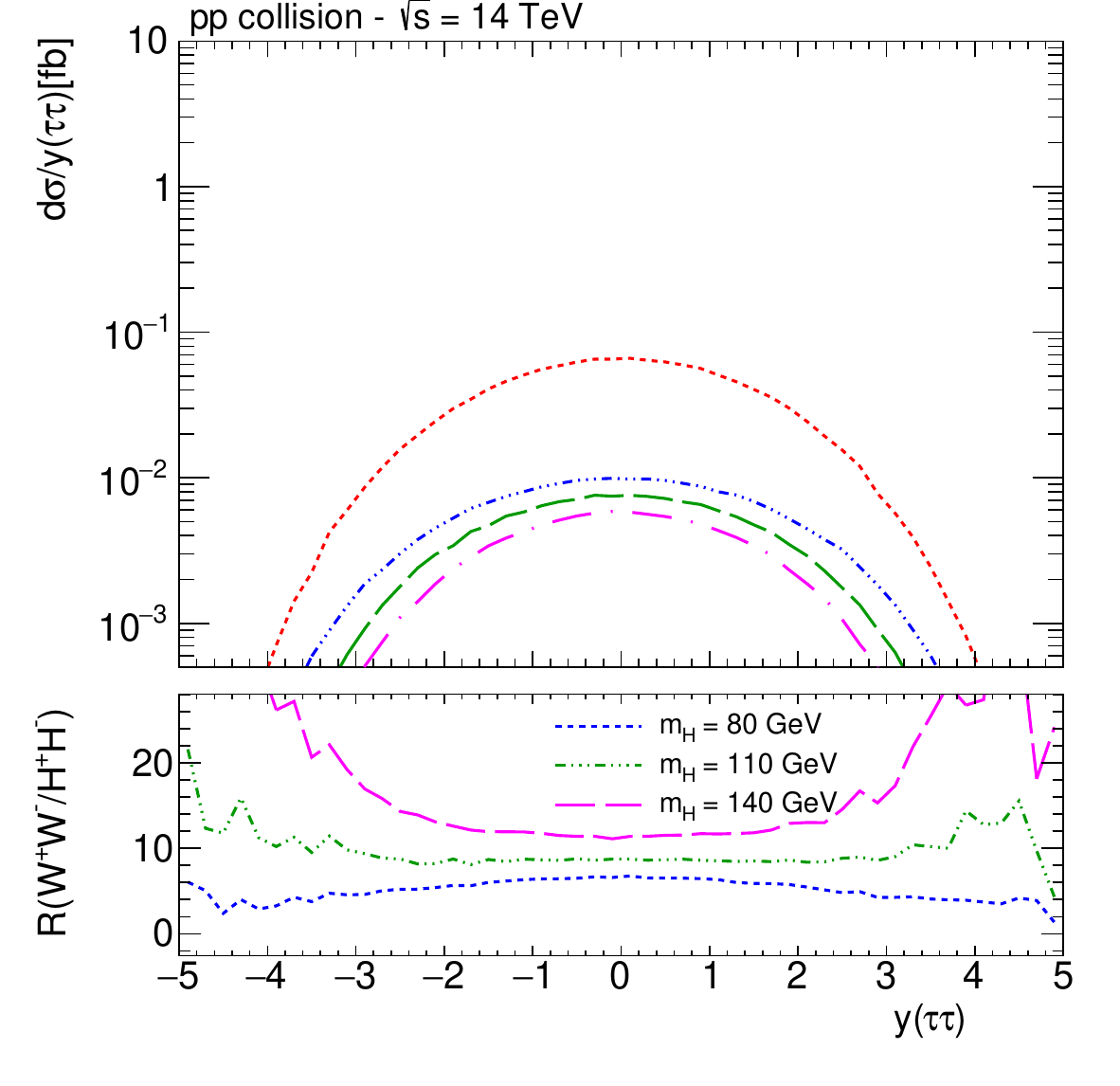}
\caption{Predictions for the transverse momentum [$p_T(\tau\tau)$], invariant mass [$m_{\tau\tau}$] and rapidity [$y(\tau\tau)$] distributions of the $\tau\tau$ pair associated with  the signal  [$pp \rightarrow p H^{+}H^{-}p \rightarrow p [\tau^{+}\nu_{\tau}][\tau^{-}\bar{\nu}_{\tau}]p$], derived assuming distinct values of  $M_{H^{\pm}}$ and considering $pp$ collisions at $\sqrt{s} = 14$ TeV. The predictions for the background 
$pp \rightarrow p \otimes  W^+ W^- \otimes p \rightarrow p \otimes  [\tau^{+}\nu_{\tau}][\tau^{-}\bar{\nu}_{\tau}] \otimes p$ are also presented.  The lower panels in the distinct plots present the ratio between the background and signal predictions.   }
\label{Fig:distr}
\end{figure}

In Fig. \ref{Fig:distr} we present our predictions for the transverse momentum [$p_T(\tau\tau)$], invariant mass [$m_{\tau\tau}$] and rapidity [$y(\tau\tau)$] distributions of the $\tau\tau$ pair associated with  the  
$pp \rightarrow p H^{+}H^{-}p \rightarrow p [\tau^{+}\nu_{\tau}][\tau^{-}\bar{\nu}_{\tau}]p$ and  $pp \rightarrow p \otimes  W^+ W^- \otimes p \rightarrow p \otimes  [\tau^{+}\nu_{\tau}][\tau^{-}\bar{\nu}_{\tau}] \otimes p$ processes. The shape of the distributions for the signal assuming different values for the single charged Higgs mass are similar, decreasing in magnitude for larger values of $M_{H^{\pm}}$.  Moreover, the position of the maximum in the $p_T(\tau\tau)$ and $m_{\tau\tau}$ distributions is dependent on the mass of the charged Higgs. Our results also indicate that the background predicts distributions that are similar to those from the signal, but with a larger normalization. 
In order to quantify the difference between the signal and background results for different values of 
$p_T(\tau\tau)$, $m_{\tau\tau}$ and $y(\tau\tau)$, we present in the lower panels of the plots shown in 
Fig. \ref{Fig:distr}, the predictions for the ratio between the distributions associated with the $W^+ W^-$
and $H^+ H^-$ production, derived assuming  distinct values of $M_{H^{\pm}}$. One has that the background dominates and is, in general, a factor $\ge 4$ than the signal.

\begin{table}[t]
\begin{center}
\scalebox{0.85}{
\begin{tabularx}{\textwidth}{|c|*{3}{Y|} c|c|}
\hline
\hline 
$pp$ @ 14 TeV & \multicolumn{3}{c|}{\bf Signal} & \multicolumn{2}{c|}{\bf Backgrounds}   \\     
\hline 
\hline
 & \multicolumn{3}{c|}{$pp \rightarrow p H^{+}H^{-}p \rightarrow p [\tau^{+}\nu_{\tau}][\tau^{-}\bar{\nu}_{\tau}]p$}   & $ pp \rightarrow p W^{+}W^{-} p \rightarrow p [\tau^{+}\nu_{\tau}][\tau^{-} \nu_{\tau}]p$ & 
 $pp \rightarrow p \tau^{-}\tau^{+} p$    \\     
    \hline
$M_{H^{\pm}}$ [GeV]&   80    & 110    & 140    &  -  &  -               \\
\hline
\hline
   $\sigma$(fb) - w/o cuts & 0.20  & 0.14  & 0.10     &1.21 &235500   \\
 \hline
$R \le 0.9$ &  0.2  & 0.14  & 0.1  & 1.21 & 0.0  \\  
    \hline  
$[1 -(\Delta \phi/\pi)] \ge 0.01$ &  0.2    & 0.14   & 0.1  & 1.19 &  0.0 \\
 \hline  
$ -2.0 \le y(\tau \tau) \le + 2.0$ & 0.16 & 0.12  & 0.086 & 1.00 &  0.0 \\
    \hline  
$ 70.0 \le m_{\tau \tau} \le 90.0$ GeV & 0.024   &0.015 & 0.008     & 0.101 & 0.0   \\
 \hline  
$ 100.0 \le m_{\tau \tau} \le 120.0$ GeV & 0.097  &   0.049 & 0.025  & 0.410  &  0.0 \\ 
\hline  
$ 130.0 \le m_{\tau \tau} \le 150.0$ GeV & 0.149  &  0.106 &   0.077   &  0.94 &  0.0 \\
    \hline      
\end{tabularx}}
\end{center}
\caption{Predictions associated with the scenario I for the total cross-sections of  single charged Higgs pair production  via photon-photon interactions in $pp$ collisions at $\sqrt{s} = 14$ TeV,  derived assuming the Type - I 2HDM, different values for the mass $M_{H^{\pm}}$. The results were estimated  considering a central detector  and kinematical cuts on the acoplanarity, ratio $R$, rapidity and invariant mass of the $\tau \tau$ pair  system. Results for the main backgrounds are also presented.} 
\label{tab:atlas}
\end{table}

In what follows, we will investigate the impact of distinct kinematical cuts on the predictions for the total cross-sections considering the two scenarios discussed above. In particular,   a cut on the ratio  $R = m_{\tau \tau}/m_X$ will be considered in the case of the  scenario I, since  the protons in the final state are assumed to be tagged, and the central mass $m_X$ can be estimated. In contrast, such a cut cannot be applied for the scenario II. Our results for the scenario I are presented in Table \ref{tab:atlas}. As expected from the results presented in Fig. \ref{Fig:acop}, the cut on $R$ is able to suppress the contribution associated with the $pp \rightarrow p \tau^{-}\tau^{+} p$ process, without impact on the other channels. The cut on the acoplanarity implies a small reduction of the $W^+ W^-$ background. On the other hand, if we impose that the $\tau^+\tau^-$ pair must be produced at central rapidities, the predictions associated with the signal and background are reduced by $\approx 17 \%$. One has that the signal predictions are of the order of $\approx 0.1$ fb, while the background one is 1 fb.
In addition, one has analyzed the impact of a cut on the invariant mass of the $\tau^+\tau^-$ pair. In particular, motivated by the results presented in Fig. \ref{Fig:distr} for the ratio between background and signal predictions for the invariant mass distributions, we have considered the selection of events in  different ranges of $m_{\tau \tau}$, where this ratio assumes its smaller values. One has that such a cut implies the reduction  of the  total cross-sections, especially if events with smaller $m_{\tau \tau}$ are selected. However, in these cases we have the larger  signal/background ratio, being $\approx 1/4$, in agreement with the results presented in Fig. \ref{Fig:distr}. Such a result indicates that, for a single charged Higgs with a small mass, the exclusive $H^+ H^-$ production gives a non - negligible contribution for the $[\tau^{+}\nu_{\tau}][\tau^{-}\bar{\nu}_{\tau}]$ final state.

\begin{table}[t]
\begin{center}
\scalebox{0.85}{
\begin{tabularx}{\textwidth}{|c|*{3}{Y|} c|c|}
\hline
\hline 
$pp$ @ 14 TeV & \multicolumn{3}{c|}{\bf Signal} & \multicolumn{2}{c|}{\bf Backgrounds}   \\     
\hline 
\hline
 & \multicolumn{3}{c|}{$pp \rightarrow p H^{+}H^{-}p \rightarrow p [\tau^{+}\nu_{\tau}][\tau^{-}\bar{\nu}_{\tau}]p$}   & $ pp \rightarrow p W^{+}W^{-} p \rightarrow p [\tau^{+}\nu_{\tau}][\tau^{-} \nu_{\tau}]p$ & 
 $pp \rightarrow p \tau^{-}\tau^{+} p$    \\     
    \hline
$M_{H^{\pm}}$ [GeV]&   80    & 110    & 140    &  -  &  -               \\
\hline
\hline
   $\sigma$(fb) - w/o cuts & 0.20   & 0.14  & 0.10     &1.21 &235500   \\
    \hline  
$[1 -(\Delta \phi/\pi)] \ge 0.01$ & 0.19   & 0.14  & 0.098     & 1.19 & 0.0  \\
    \hline  
$ 2.0 \le y(\tau \tau) \le  4.5$ & 0.019   & 0.011   &  0.006    & 0.093 & 0.0  \\
    \hline  
$ 70.0 \le m_{\tau \tau} \le 90.0$ GeV & 0.003   & 0.0015  &  0.0007   & 0.012 & 0.0  \\
 \hline  
$ 100.0 \le m_{\tau \tau} \le 120.0$ GeV &  0.002  &0.0008 &  0.0006    & 0.009 & 0.0  \\ 
\hline  
$ 130.0 \le m_{\tau \tau} \le 150.0$ GeV & 0.001   &  0.0009 & 0.0006     & 0.006 & 0.0  \\
    \hline      
\end{tabularx}}
\end{center}
\caption{Predictions associated with the scenario II for the total cross-sections of single charged Higgs pair production  via photon-photon interactions in $pp$ collisions at $\sqrt{s} = 14$ TeV,  derived assuming the Type - I 2HDM, different values for the mass $M_{H^{\pm}}$. The results were estimated considering a forward detector  and kinematical cuts on the acoplanarity,  rapidity and invariant mass of the $\tau \tau$ pair  system. Results for the main backgrounds are also presented.} 
\label{tab:lhcb}
\end{table}

In Table \ref{tab:lhcb} we present the results associated with the scenario II, which can be investigated by the LHCb Collaboration. As explained before, for this scenario, the cut on $R$ cannot be applied, since the central mass $m_X$ cannot be reconstructed without the tagging of the protons in the final state. However, the results presented in Table \ref{tab:lhcb} indicate that the cut on the acoplanarity is able to fully suppress the contribution associated with the $pp \rightarrow p \tau^{-}\tau^{+} p$ process. The selection of forward events implies the reduction of the cross - sections by almost one order of magnitude. Moreover, the cut on the invariant mass suppress the predictions by a factor $\ge 8$, implying that the cross - sections become of the order of $10^{-3}$ fb, making a future experimental analysis of the exclusive $H^+ H^-$ production at forward rapidities a hard task.

\section{Summary and concluding remarks}
\label{sec:sum}
Over the last decades, the study of photon - induced interactions in hadronic colliders became a reality, such that currently the Large Hadron Collider (LHC) is also considered a powerful photon - photon collider, which can be used to improve our understanding of the Standard Model as well as to searching for New Physics.  The current data have already constrained several BSM scenarios, and more precise measurements are expected in the forthcoming years. Such an expectation has motivated the exploratory study performed in this paper, where we have considered the exclusive single charged Higgs pair production in $pp$ collisions at $\sqrt{s} = 14$ TeV. One has assumed the type - I two - Higgs-Doublet model, which  is one of the simplest BSM frameworks that predict charged Higgs bosons and that allows light charged Higgs with mass below 100 GeV. Our study complements the analysis performed in Ref. \cite{Cheung:2022ndq}, where the single charged Higgs pair production in inelastic $pp$ collisions was estimated and the current viable parameters for the type - I 2HDM were derived.      The exclusive $H^+ H^-$ production cross - section was estimated considering different values for the charged Higgs and two distinct experimental configurations, whose are similar to those present in central (ATLAS/CMS) and forward (LHCb) detectors. We focused on the leptonic $H^{\pm}\rightarrow [\tau\nu_{\tau}]$ decay mode, which one has verified to be the final state with larger signal/background ratio. We have demonstrated that the background associated to the   
 $pp \rightarrow p \tau^{-}\tau^{+} p$  can be fully removed by assuming a cut on the ratio $R = m_{\tau \tau}/m_X$ and/or in the acoplanarity. In contrast, the contribution of the 
$ pp \rightarrow p W^{+}W^{-} p \rightarrow p [\tau^{+}\nu_{\tau}][\tau^{-} \nu_{\tau}]p$ background process dominates and generates distributions that are similar those predicted by the signal. However, the signal/background ratio is of the order of 1/4 for a light charged Higgs, which implies a non - negligible contribution for the $[\tau^{+}\nu_{\tau}][\tau^{-} \nu_{\tau}]$ final state. Such a promising result motivates the extension of this exploratory study by considering more sophisticated separation methods, as e.g. those used by the experimental collaborations in Refs. \cite{ATLAS:2018gfm,CMS:2019bfg}, where kinematic variables that differentiate between the signal and backgrounds are identified and combined into a multivariate discriminant, with the output score of the boosted decision tree (BDT) used in order to separate the single charged signal from the SM background processes. We plan to perform such an extension in a forthcoming study.

\begin{acknowledgements}
D.E.M. thanks dearly, in the person of Janusz Chwastowski, the total support of the Henryk Niewodniczanski Institute of Nuclear Physics Polish Academy of Sciences (grant no. UMO2021/43/P/ST2/02279). T.B.M. acknowledges ANID-Chile grant FONDECYT No. 3220454  for financial support. This work was  partially financed by the Brazilian funding
agencies CNPq,   FAPERGS and INCT-FNA (processes number 
464898/2014-5).
\end{acknowledgements}

\end{document}